\documentclass[fleqn,usenatbib]{mnras}
\usepackage{mathptmx}
\usepackage[varvw]{newtxmath}  
\usepackage[T1]{fontenc}
\usepackage{ae,aecompl}
\usepackage{times}
\usepackage{amsmath}
\usepackage{upgreek}
\usepackage{xcolor}
\usepackage{graphicx}
\usepackage{pdflscape}
\usepackage{multicol} 
\usepackage{amsmath,bm}
\usepackage{caption}

\begin{document}

\title[Comparability of Molecular Cloud Physical Properties]{Connecting Star Formation in the Milky Way and Nearby Galaxies. I. Comparability of Molecular Cloud Physical Properties}

\author[J. W. Zhou]{
J. W. Zhou \thanks{E-mail: jwzhou@mpifr-bonn.mpg.de}$^{1}$
Sami Dib $^{2}$
\\
$^{1}$Max-Planck-Institut f\"{u}r Radioastronomie, Auf dem H\"{u}gel 69, 53121 Bonn, Germany\\
$^{2}$Max-Planck-Institut f\"{u}r Astronomie, K\"{o}nigstuhl 17, 69117 Heidelberg, Germany
}

\date{Accepted XXX. Received YYY; in original form ZZZ}
\pubyear{2024}
\maketitle

\begin{abstract}
We used CO (2–1) and CO (1–0) data cubes to identify molecular clouds and study their kinematics and dynamics in three nearby galaxies and the inner Milky Way. When observed at similar spatial and velocity resolutions, molecular clouds in the same mass range across these galaxies show broadly comparable physical properties and similar star formation rates (SFRs).
However, this comparability depends on smoothing Milky Way clouds to match the resolution of the extragalactic observations. The beam effect can artificially inflate cloud sizes, leading to inaccurate estimates of radius, density, and virial parameters. By comparing high-resolution and smoothed Milky Way data, we established criteria to exclude beam-affected clouds in the extragalactic sample.
After applying this filter, cloud properties remain consistent across galaxies, though some clouds in NGC 5236 show elevated velocity dispersions, likely due to environmental effects.
In the inner Milky Way, molecular clouds fall into two groups: those with clumps and those without. Clump-associated clouds are more massive, denser, have higher velocity dispersions, lower virial parameters, and stronger 8$\mu$m emission, suggesting more intense feedback.
Strong correlations are found between cloud mass and total clump mass, clump number, and the mass of the most massive clump. These results suggest that a cloud’s physical conditions regulate its internal clump properties and, in turn, its star-forming potential.
\end{abstract}


\maketitle

\section{Introduction}\label{sec:intro}

Galaxies function as stellar nurseries in the universe, forming stars through the gravitational collapse of the densest parts of molecular clouds within the interstellar medium (ISM). These molecular clouds are widespread throughout galaxies and mark the regions where star formation is actively taking place. Within each cloud, multiple dense clumps are typically present—these clumps are the immediate sites of star formation and serve as the progenitors of embedded star clusters
\citep{Kennicutt2012-50,Miville2017-834,Rosolowsky2021-502,Urquhart2022-510,Yan2017-607,Schinnerer2024-62,Zhou2024PASP-1,Zhou2024PASP-2}.
As presented in \citet{Motte2018-56, Vazquez2019-490,Kumar2020-642, Henshaw2020-4, Dib2020-642, Zhou2022-514,Zhou2023-676,Zhou2024-686-146, Zhou2024PASA, Zhou2024-534,Zhou2025-537b,Zhou2025arXiv250616664Z},
from galactic scales down to dense cores, the molecular gas exhibits a multi-scale, hierarchical structure characterized by hub-filament networks. Dense cores act as hubs within clumps, clumps serve as hubs within molecular clouds, and molecular clouds themselves can function as hubs within galaxies. These hubs, as local centers of gravity, can continuously accrete diffuse surrounding gas, leading to sustained mass growth. 

A thorough understanding of the properties of giant molecular clouds (GMCs) is key to deciphering the connection between gas dynamics and star formation in galaxies. Thanks to high-resolution, multi-wavelength observations, it is now possible to resolve star formation down to the scale of individual molecular clouds in nearby galaxies. In particular, high-resolution CO imaging from ALMA (Atacama Large Millimeter/submillimeter Array), along with other submillimeter observations, enables systematic investigations of the molecular cloud population beyond the Milky Way
\citep{Leroy2021-257,Leroy2021-255,Lee2022-258,Emsellem2022-659,Lee2023-944,Grishunin2024-682,Schinnerer2024-62}.

Since molecular clouds in the Milky Way can be resolved down to their internal structures, they provide a crucial reference for understanding the internal composition and star formation processes of molecular clouds in external galaxies. 
Star formation occurs primarily within clumps, the dense substructures of molecular clouds.
Star formation observed on molecular cloud scales or larger is essentially an integrated outcome of the activity occurring within these clumps. 
To truly understand the underlying physics of star formation at those larger scales, it is essential to characterize the star-forming states of individual clumps. 
This requires turning to the Milky Way as a reference, leveraging surveys such as ATLASGAL and Hi-GAL \citep{Schuller2009-504,Urquhart2022-510,Molinari2010,Elia2021-504}, which are capable of resolving individual clumps, along with numerous ALMA follow-up studies that zoom in on single clumps in detail \citep{Sanhueza2019-886,Liu2020,Motte2022-662,Molinari2025-696}. 

If we aim to use molecular clouds in the Milky Way as a benchmark, we must first establish their comparability to those in nearby galaxies. In this work, we first identify and characterize molecular clouds in the Milky Way, then conduct a detailed comparison with clouds in nearby galaxies. We emphasize the importance of matching angular resolution—i.e., smoothing the data to a common resolution—when comparing cloud properties across different observational datasets. Finally, by taking advantage of the high-resolution insights available for Galactic molecular clouds, we explore how such internal structural information can help interpret the large-scale phenomena observed in $\sim$100 pc molecular clouds in nearby galaxies through ALMA observations, and relate them to the underlying small-scale physics.


\section{Data}

\subsection{CO (2-1) cubes and mid-infrared images of nearby galaxies} 

\begin{figure*}
\centering
\includegraphics[width=1\textwidth]{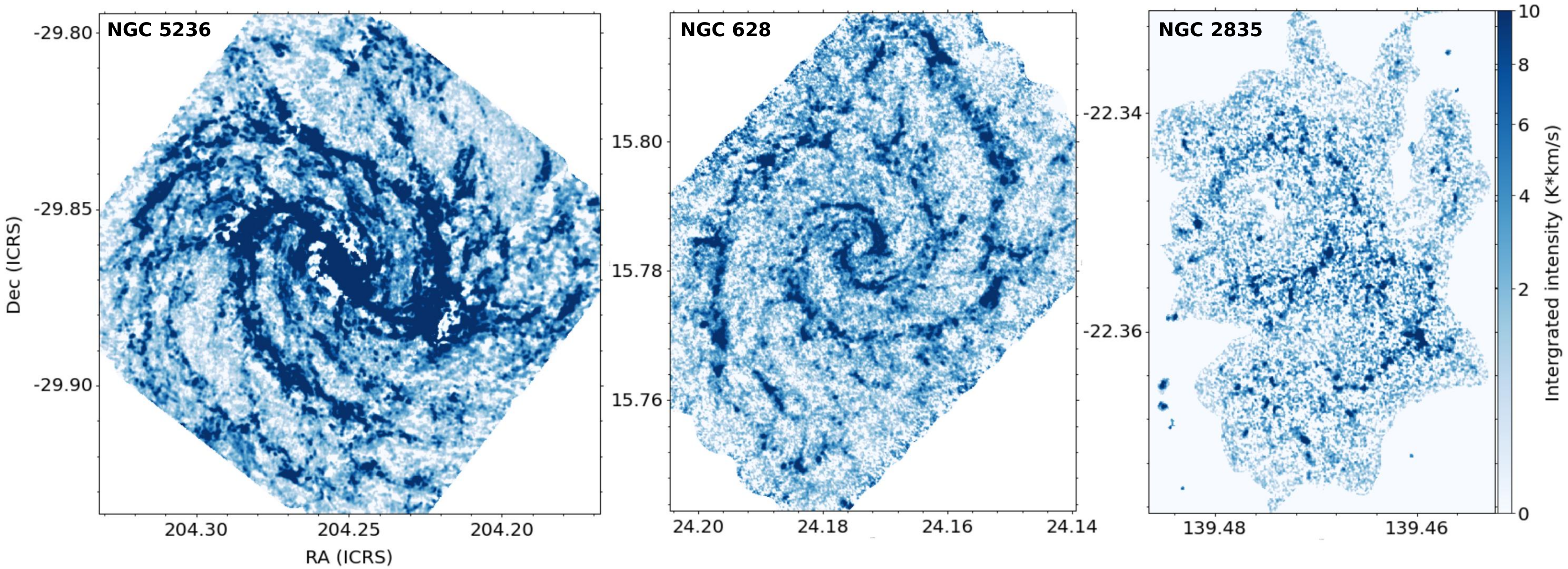}
\caption{Velocity-integrated CO (2-1) intensity maps of three nearby galaxies.}
\label{sample}
\end{figure*}

\begin{table*}
\centering
\caption{Physical parameters of three nearby galaxies—from left to right: inclination, distance, beam size, linear resolution, pixel size , beam area in pixels and number of molecular clouds.}
\begin{tabular}{cccccccc}
\hline
	&$i$ (deg) &D (Mpc) &beam size (")&linear (pc)&pixel (")&beam area (pix)&leaves\\
NGC 5236	&	24	&4.89	&	2.14	&50.7	 &0.5&21&1722 \\
NGC 628	&	8.9	 &9.84	&	1.12	&53.5	 &0.2&36&773 \\
NGC 2835	&	41.3	&12.22	&	0.84	&50	 &0.2&20&130 \\
MW-North & &&72&&24&&597\\
MW-South & &&72&&24&&707\\
\hline
\label{near}
\end{tabular}
\end{table*}

As shown in Table.\ref{near},
we selected three nearby galaxies (NGC 5236, NGC 628 and NGC 2835) with a linear resolution $\sim$50 pc from the PHANGS-ALMA survey \citep{Leroy2021-257,Leroy2021-255}, which show different morphologies in Fig.\ref{sample}.
We used the combined 12m+7m+TP CO (2$-$1) data cubes to investigate gas kinematics and dynamics, which have a spectral resolution of 2.5 km s$^{-1}$. In the Milky Way, the typical molecular cloud size is $\sim$30 pc \citep{Miville2017-834}. For molecular clouds smaller than 50 pc in the three nearby galaxies, observational results are inevitably affected by the beam effect.

Among the three galaxies, only NGC 2835 and NGC 628 have available James Webb Space Telescope (JWST) images from the PHANGS-JWST survey \citep{Lee2023-944}. We only use the 21 $\mu$m images
in this work.

\subsection{CO (1-0) cubes and mid-infrared images of the inner Milky Way}

Table.1 of \citet{Schuller2021-500} summarizes the CO surveys in the Milky Way. Since we will study the gas environment around the ATLASGAL clumps ($|l| < 60^\circ$ and $|b| < 1.5^\circ$) \citep{Urquhart2018-473},
we focus on CO surveys that cover a wide range of Galactic latitudes. 
In this work, we use the $^{12}$CO/$^{13}$CO (1-0) cubes from the ThrUMMS (Three-mm Ultimate Mopra Milky Way Survey)  and FUGIN (FOREST Unbiased Galactic plane Imaging survey with the
Nobeyama 45 m telescope) surveys. The observations and data reduction of the two surveys have been described in detail in
\citet{Barnes2015-812, Barnes2025arXiv, Umemoto2017-69}.

The ThrUMMS survey, with an angular resolution of $72\arcsec$ and a spectral resolution of 0.34 km\,s$^{-1}$, covers the Galactic plane between $300^\circ \le l \le 360^\circ$ and $-1^\circ \le b \le 1^\circ$. For the FUGIN survey, which has an angular resolution of $20\arcsec$ and a spectral resolution of 1.3 km\,s$^{-1}$, we focus on observations within the inner Milky Way, covering the Galactic plane between $10^\circ \le l \le 50^\circ$ and $-1^\circ \le b \le 1^\circ$. To compare with the selected nearby galaxies, we unify these two datasets to an angular resolution of 72$\arcsec$ with a pixel size of 24$\arcsec$ and a spectral resolution of 2.5 km s$^{-1}$ with a velocity channel of 1.25 km s$^{-1}$.

We also use the mid-infrared images from the {\it Spitzer} archive, i.e. 8 and 24 $\mu$m images from the GLIMPSE \citep{Benjamin2003-115} and MIPSGAL \citep{Carey2009-121} surveys, respectively.

\section{Methodology}

\subsection{Cloud identification}

\subsubsection{Nearby galaxies}

The cloud identification for NGC 5236 and NGC 628 has been done in \citet{Zhou2024PASA,Zhou2024-534}.
We conducted a direct identification of hierarchical (sub-)structures based on the 2D intensity maps. As described in \citet{Rosolowsky2008-679}, the dendrogram algorithm decomposes density or intensity data into hierarchical structures called leaves, branches, and trunks. 
In the {\it astrodendro} package \footnote{\url{https://dendrograms.readthedocs.io/en/stable/index.html}},
there are three major input parameters for the dendrogram algorithm: {\it min\_value} for the minimum value to be considered in the dataset, {\it min\_delta} for a leaf that can be considered as an independent entity, and {\it min\_npix} for the minimum area of a structure.

Given that all retained structures in the strictly masked Moment 0 map\footnote{Details of the masking strategy and completeness statistics are provided in the PHANGS pipeline paper \citep{Leroy2021-255}.} are considered reliable, we only require that the smallest identified structure exceeds one beam area (see Table.\ref{near}). No additional constraints are imposed within the algorithm to reduce the dependence of the identification results on parameter choices. The number of identified molecular clouds (leaves) in each galaxy is shown in Table.\ref{near}.

\subsubsection{The inner Milky Way}\label{inner}

\begin{figure*}
\centering
\includegraphics[width=1\textwidth]{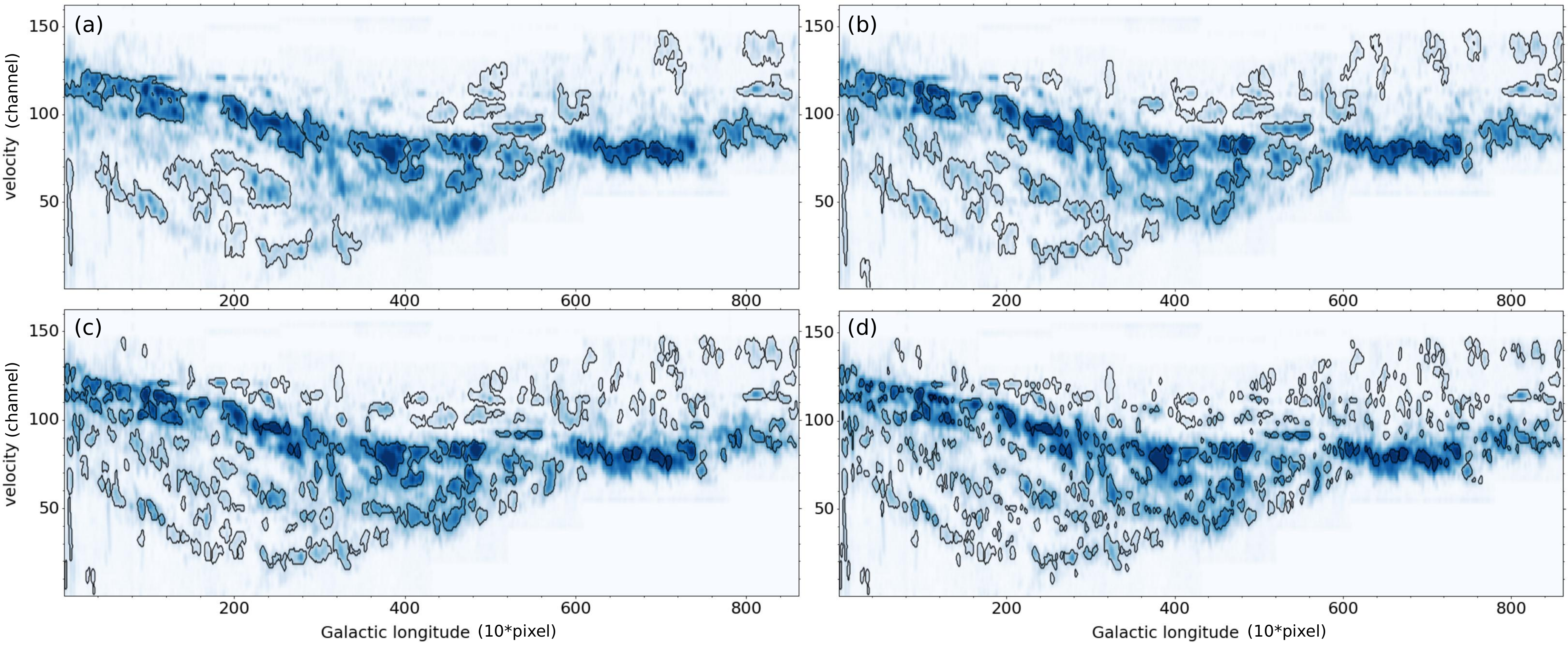}
\caption{The Galactic longitude–velocity map of the Milky Way with $300\degr\le l \le 360\degr$ and $-1\degr \le b \le 1\degr$ from the ThrUMMS survey. Black contours show the clouds identified by the dendrogram algorithm with different {\it min\_npix} values (250, 100, 25 and 5 pixels). Here, both the velocity and spatial axes are expressed in pixel units.}
\label{lv}
\end{figure*}


To enable comparison with molecular clouds in nearby galaxies with the resolution of $\sim$50 pc, it is important to identify cloud structures on large spatial scales. 
Molecular clouds often appear as complexes or as couplings of multi-scale hub-filament structures \citep{Zhou2025arXiv250616664Z}, containing numerous substructures within. Our goal is to identify molecular clouds that are as spatially and kinematically continuous and extended as possible. 
Since the observations for the molecular gas are confined to within
$\pm$1$^{\rm o}$ in Galactic latitude ($b$), we can directly identify molecular clouds from the Galactic longitude–velocity ($lv$) map, as shown in Fig.\ref{lv}. For face-on nearby galaxies, we identify molecular clouds directly on the galactic disk, where the clouds are spatially separated. In the Milky Way, however, molecular clouds overlap along the line of sight. By identifying molecular clouds in the $lv$ map, we can use the velocity dimension to separate them. Considering the hierarchical structure of molecular gas, we identify structures at different scales and finally remove duplicate structures.

Before exporting the $lv$ map, we already applied a 3*rms mask to the $lbv$ cube. As described above, we do not consider the value settings of {\it min\_value} and {\it min\_npix} in the dendrogram algorithm. 
Fig.\ref{lv} shows the structures identified with different {\it min\_npix} values. Here, 
both the velocity and spatial axes are expressed in pixel units, so that they can be properly matched.
Finally, we take {\it min\_npix} = 500, 250, 100, 50, 25, 10, and 5 pixels, in order to cover continuous structures across different scales in the $lv$ map.
Since we used a set of different {\it min\_npix} values, each {\it min\_npix} produces a set of structures. For two adjacent {\it min\_npix} values, which are numerically close, they tend to identify a large number of the same structures (i.e., duplicate structures).
To exclude duplicate structures,
for two structures with similar central coordinates (with areas $A_{\rm 1}$ and $A_{\rm 2}$, where $A_{\rm 1}$ > $A_{\rm 2}$), if ($A_{\rm 1}$ - $A_{\rm 2}$)/$A_{\rm 2}$ < 0.3, the structure with area $A_{\rm 2}$ is removed. Because our goal is to eliminate duplicate identifications rather than nested structures, we adopt a value of 0.3.
Finally, we identify 707 and 597 molecular clouds from the ThrUMMS and FUGIN surveys, respectively.

We derived the $lv$ map from the $lbv$ cube. Clouds were identified using this $lv$ map, which provides the Galactic longitude, velocity, and their ranges ($l_0$, $v_0$, $dl$, and $dv$) for each cloud, while the Galactic latitude and its range ($b_0$ and $db$) remain unknown. Observations already limit the Galactic latitude to $\pm 1^\circ$. Using $(l_0, 0^\circ, v_0)$ as the center and $(dl, 2^\circ, dv)$ as the extraction range, we extract the molecular gas from the $lbv$ cube. $b_0$ is then determined from the region of strongest emission within this map. Then we take $db = dl$ and use $(l_0, b_0, v_0)$ as the center and $(dl, dl, dv)$ as the extraction range to extract the molecular gas from the $lbv$ cube again to further fix the value $db$.



\citet{Urquhart2022-510} and \citet{Duarte2021-500} have provided detailed descriptions of strategies for determining the distances to molecular clouds and clumps in the inner Milky Way.
In order to determine the distances of the identified molecular clouds and investigate the correlations between the physical properties of clouds and those of their internal clumps,
we first match the ATLASGAL clumps \citep{Urquhart2022-510} to the identified clouds based on their spatial and velocity ranges, and then assign the median distance of all associated clumps as the distance of the molecular cloud. A typical velocity difference between the median velocity of all associated clumps and the cloud velocity is $\sim$1.2 km/s. 
For molecular clouds without matched clumps, we cross-match them with the existing molecular cloud catalogs derived from the SEDIGISM (Structure, Excitation, and Dynamics of the Inner Galactic InterStellar
Medium) survey \citep{Duarte2021-500} to determine their distances. 
Here, we require the spatial positions of the molecular clouds in the two catalogs to be as close as possible (i.e., we adopt the pair with the smallest separation in both spatial distance and velocity). We also require that the velocity difference of the pair is less than 10 km/s and that the spatial distance of the pair is smaller than the cloud size.
As a result, a typical velocity difference between the closest cloud pair (one from our cloud catalog and the other from \citep{Duarte2021-500}) is $\sim$1.0 km/s. 
Finally, for the ThrUMMS survey, 570 and 137 molecular clouds have and do not have distance estimates, respectively. For the FUGIN survey, the numbers of molecular clouds with and without distance estimates are 370 and 227, respectively. 
Since the SEDIGISM survey is restricted to $|b| < 0.5^\circ$,
therefore, a significant fraction of molecular clouds in the ThrUMMS and FUGIN surveys are not covered by the SEDIGISM survey.
In the following analysis, we focus only on molecular clouds with distance estimates.


\subsection{Physical parameters}\label{parameters}

For molecular clouds identified in the Milky Way, the intensity-weighted velocity dispersion is calculated by the second moment method,
\begin{equation}
v_0 = \frac{\sum T(v) * v}{\sum T(v)}, \quad
\sigma = \sqrt{ \frac{ \sum T(v) * (v - v_0)^2 }{ \sum T(v) } }.
\label{mt}
\end{equation}
For molecular clouds in nearby galaxies, we perform Gaussian fitting on the CO line profile to determine the velocity range of the emission. This step is necessary for the velocity dispersion calculation by the second moment.
All calculations are based on the PPV cubes masked at the 3*rms level.
The average column density of a cloud is,
\begin{equation}
  N  =  I_{\rm CO}^{\rm tot} * X_{\rm CO}/n_{\rm pix},
\label{N-o}
\end{equation}
where $I_{\rm CO}^{\rm tot}$ is the total velocity-integrated intensity of the cloud. $n_{\rm pix}$ is the total number of non-empty pixels. The $X_{\rm CO}$ is taken to be $ 2 \times 10^{20}~\mathrm{cm}^{-2}~(\mathrm{K~km~s}^{-1})^{-1}$ \citep{Bolatto2013-51}.
The total area of the cloud is,
\begin{equation}
  A = n_{\rm pix} * l_{\rm pix}^2,
\end{equation}
where $l_{\rm pix}$ is the length of a pixel.
The total mass of the cloud is,
\begin{equation}
  M_{\rm cloud} = N * \mu_{\rm H_{2}} * m_{\rm H} * A,
\end{equation}
where $\mu_{\rm H_{2}} = 2.8$ is the molecular weight per hydrogen molecule, 
$m_{\rm H}$ is the hydrogen atom mass.
The minimum and maximum radii of the cloud are,
\begin{equation}
  r_{\rm min} = \sqrt{A/\pi} 
\end{equation}
and
\begin{equation}
  r_{\rm max} = \sqrt{n_{\rm x}*n_{\rm y}}*l_{\rm pix} ,
  \label{rmax}
\end{equation}
where $n_{\rm x}$ and $n_{\rm y}$ are the length and width of the box enclosing the cloud. 
The average column density of the cloud after smoothing is,
\begin{equation}
  N_{\rm a} = M/(\pi*r_{\rm s}^2)/(\mu_{\rm H_{2}} * m_{\rm H}),
\label{N-a}
\end{equation}
where $r_{\rm s}$ is the radius of the cloud after smoothing.
The virial parameter of the cloud after smoothing is,
\begin{equation}
  \alpha_{\rm vir} = 5*\sigma^2*r_{\rm s}/(G*M).
\end{equation}

\subsection{Radius after smoothing}\label{smooth}

\begin{figure*}
\centering
\includegraphics[width=1\textwidth]{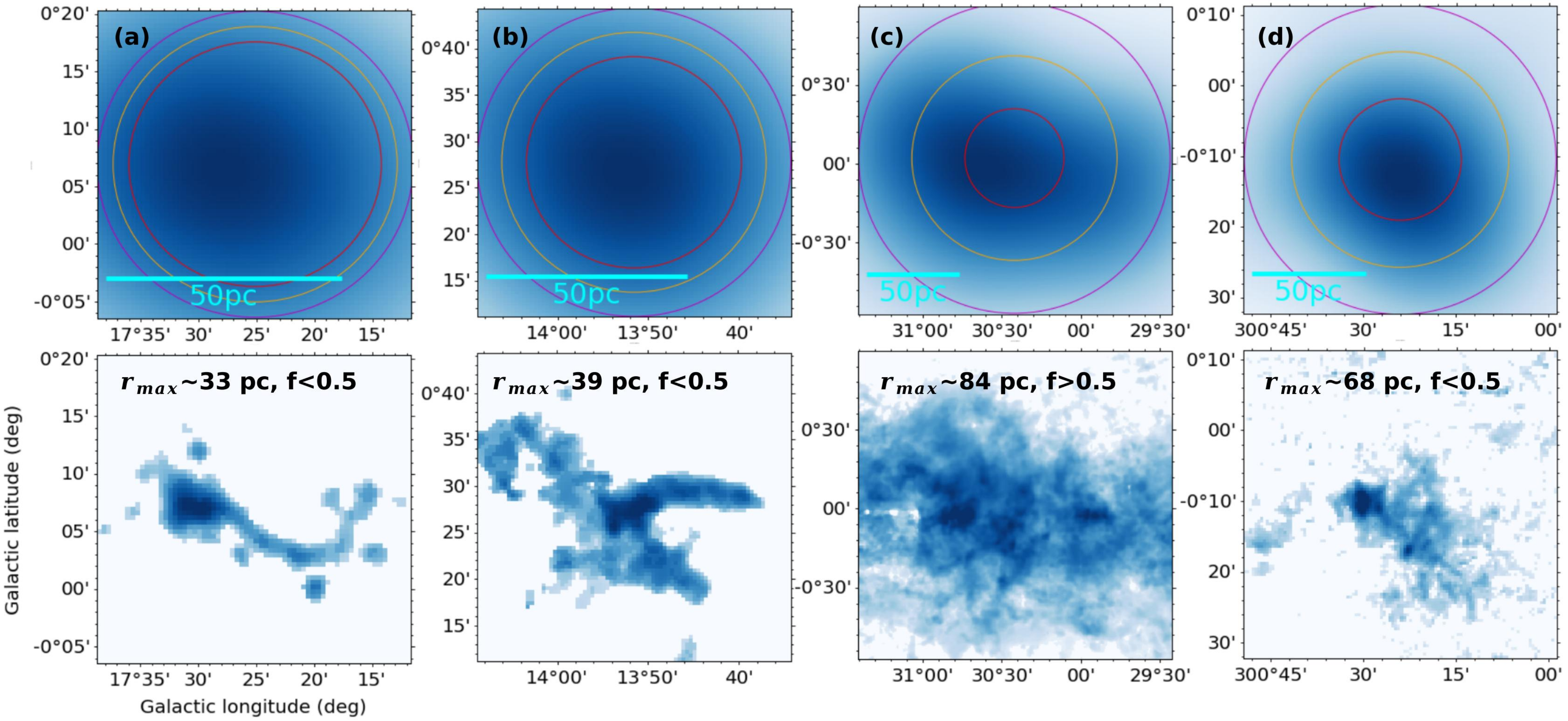}
\caption{The first and second rows show the smoothed and original molecular clouds, respectively. In the first row, the three concentric circles represent three different radius estimates, i.e. 26.75 pc, ($r_{\rm max}$+26.75 pc)/2 and $r_{\rm max}$, from innermost to outermost. $f$ is the filling factor. $r_{\rm max}$ is defined in Sec.\ref{parameters}.}
\label{case}
\end{figure*}

In order to compare with molecular clouds in nearby galaxies, we need to smooth the high-resolution molecular clouds in the Milky Way to match the resolution of extragalactic observations. This step significantly affects the estimation of the cloud radius.
This also leads to notable changes in the estimates of other physical quantities, as shown in Sec.\ref{parameters}.
In Table.\ref{near}, the linear resolution of NGC 628 is 53.5 pc.
By taking into account the distances of the identified clouds in the Milky Way, we adjust the smoothing factor such that all clouds are brought to a common spatial resolution of 53.5 pc.
Since we are to compare with individual molecular clouds identified in nearby galaxies, for the smoothed map, we only consider the map with one main emission structure, as shown in Fig.\ref{case}. 

When we smooth the high-resolution image to the lower-resolution one, the radii of structures in the original high-resolution image increase due to blurring on one hand, and on the other hand, spatially dispersed structures may merge into a larger structure. Therefore, the radius of a structure after smoothing depends not only on its original size, but also on its spatial distribution in the original image.
Here, we define the filling factor ($f$) as the ratio of pixels with emission to the total number of pixels in the image.  
The radii of the smoothed molecular clouds can generally be divided into four categories:
(1) For the clouds with $r_{\rm max}$ < 26.75 pc (half the beam size), 
the smoothed structure is fully contained within a single beam, so its size is effectively the beam size after smoothing, i.e. $r_{\rm s}$ = 26.75 pc;
(2) For the clouds with 26.75 < $r_{\rm max}$ < 53.5 pc, after smoothing, the emission typically fills the entire image, and thus $r_{\rm s}$ = $r_{\rm max}$ (Fig.\ref{case}(a) and (b));
(3) For the clouds with $r_{\rm max}$ > 53.5 pc and $f$ > 0.5, the emission also typically fills the entire image after smoothing, leading to $r_{\rm s}$ = $r_{\rm max}$ (Fig.\ref{case}(c));
(4) For the clouds with $r_{\rm max}$ > 53.5 pc and $f$ < 0.5, their radii typically range between 26.75 pc and $r_{\rm max}$, so we adopt a radius of ($r_{\rm max}$ + 26.75 pc) / 2 (Fig.\ref{case}(d)).

\subsection{Star formation rates}

We further calculate the star formation rates (SFRs) of molecular clouds in the Milky Way and nearby galaxies. For nearby galaxies, JWST observations can resolve individual molecular clouds. 
In \citet{Zhou2024-534}, for the clouds in NGC 628, we calculated the SFR surface density of a cloud using the JWST 21 $\mu$m image following the prescriptions described in \citet{Leroy2021-257}, 
\begin{equation}
    \frac{\Sigma_{\rm SFR}}{\rm M_{\odot}~yr^{-1}~kpc^{-2}} = 3.8\times10^{-3} \left(\frac{I_{\rm 21 \mu m}}{\rm MJy~sr^{-1}}\right),
\end{equation}
\begin{equation}
    \mathrm{SFR} = \Sigma_{\rm SFR}*l_{\rm pix}^2
    \label{21um}
\end{equation}
where $I_{\rm 21 \mu m}$ is the total 21 $\mu$m flux contained within the cloud. 

For the clouds in the Milky Way, we also use equation.\ref{21um} to calculate the SFR surface density, 
but we replace the JWST 21 $\mu$m image by the {\it Spitzer} 24 $\mu$m image. 
Unlike in face-on nearby galaxies, 
24 $\mu$m emission in the Milky Way suffers from severe overlap due to projection effects. On galaxy-wide and cloud scales, the SFR is correlated with the total amount of dense molecular gas \citep{Gao2004-606,Lada2010-724,Shimajiri2017-604}. For each molecular cloud, we know its spatial extent. By calculating the total velocity-integrated CO intensity within this region ($I_{\rm CO,tot}$) and the portion of that intensity confined to the velocity range of the cloud ($I_{\rm CO,cloud}$), we obtain a scaling factor, i.e. $f_{24 \mu m}= I_{\rm CO,cloud}/I_{\rm CO,tot}$. We then multiply the total 24 $\mu$m flux within the region by this factor to determine the 24 $\mu$m flux associated with the cloud. As shown in Fig.\ref{sfr-m}, this method proves to be effective.


\section{Results and discussion}

\subsection{Comparability of physical properties}\label{compare}

\begin{figure*}
\centering
\includegraphics[width=1\textwidth]{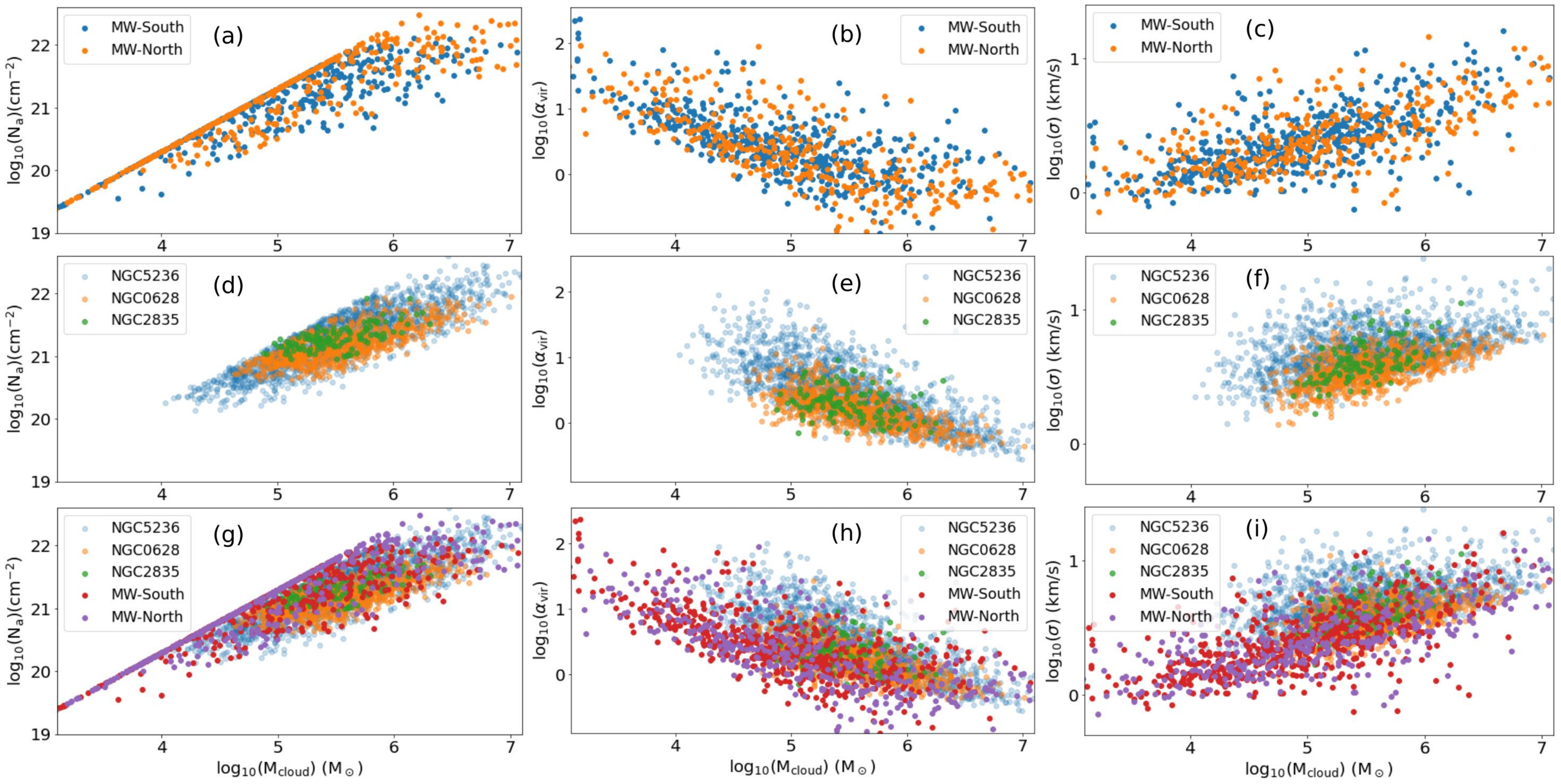}
\caption{Comparison of the physical properties of molecular clouds in the southern and northern inner Milky Way (first row), in three nearby galaxies (second row), and between molecular clouds in the inner Milky Way and those in nearby galaxies (third row).
$M_{\rm cloud}$, $\sigma$, $N_{\rm a}$ and $\alpha_{\rm vir}$ represent the cloud mass, velocity dispersion, average column density, and virial parameter, respectively.}
\label{para}
\end{figure*}

\begin{figure}
\centering
\includegraphics[width=0.4\textwidth]{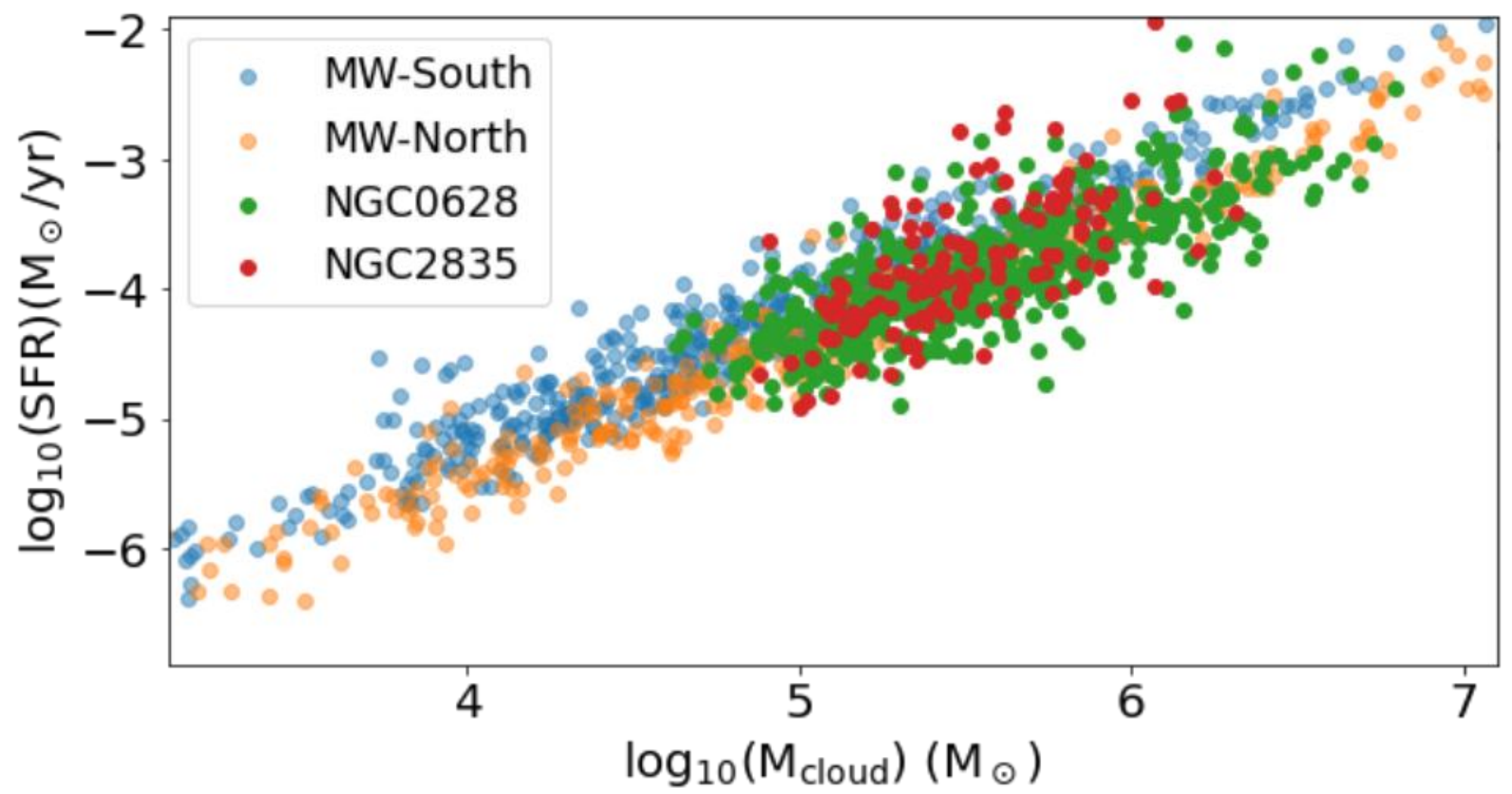}
\caption{Comparison of star formation rates (SFRs) in molecular clouds within the inner Milky Way and in nearby galaxies.}
\label{sfr-m}
\end{figure}

As shown in Fig.\ref{para}, we compare molecular clouds in the southern and northern inner Milky Way, those in three nearby galaxies, and molecular clouds in the inner Milky Way and nearby galaxies. As long as the observational resolution across these galaxies is consistent, the physical properties of their molecular clouds are overall comparable. In Fig.\ref{sfr-m}, the SFRs of molecular clouds in the inner Milky Way and nearby galaxies are also comparable, consistent with the similarity in their physical properties. Moreover, molecular cloud mass shows significant correlations with other physical parameters.


In the upper right boundary of Fig.\ref{para}(a), the average column density increases linearly with mass, which implies a constant radius. These molecular clouds belong to the category with $r_{\rm max} < 26.75$ pc, as defined in Sec.\ref{smooth}. In Fig.\ref{para}(g), this linear increasing boundary aligns well with observations in nearby galaxies. This further highlights the importance of using a consistent resolution when comparing observations of the Milky Way with those of nearby galaxies.

\subsection{Original and smoothed molecular clouds}\label{original}

\begin{figure*}
\centering
\includegraphics[width=1\textwidth]{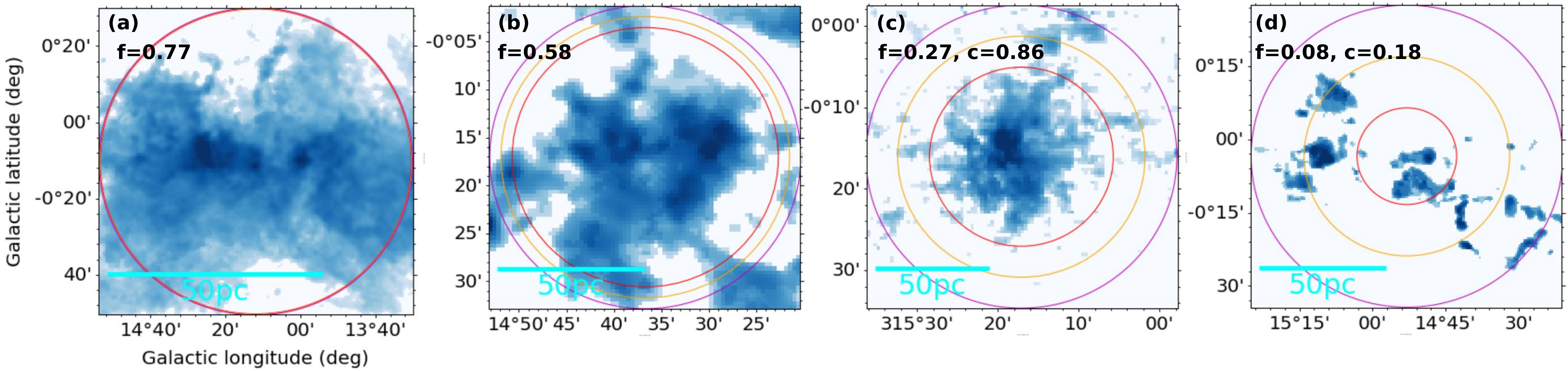}
\caption{Estimation of the true radius of the original molecular clouds. The three concentric circles, from innermost to outermost, represent three different radius estimates, i.e. $r_{\rm min}$, ($r_{\rm max}$+$r_{\rm min}$)/2 and $r_{\rm max}$, respectively. $f$ is the filling factor. $r_{\rm max}$ is defined in Sec.\ref{original}. In panel (a), due to the high filling factor, $r_{\rm max} \sim r_{\rm min}$.}
\label{true}
\end{figure*}

\begin{figure*}
\centering
\includegraphics[width=0.75\textwidth]{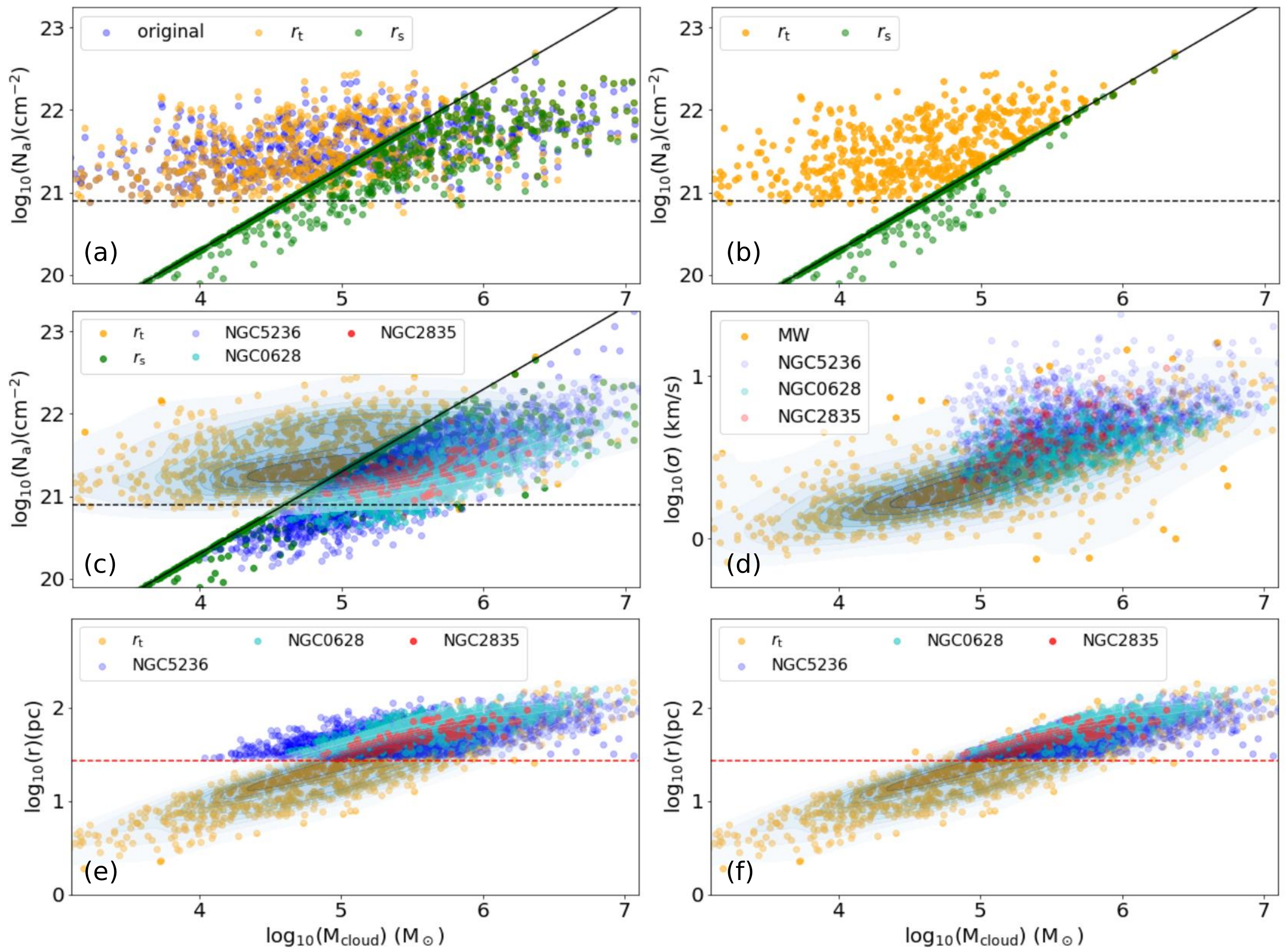}
\caption{The impact of smoothing on the physical properties of molecular clouds. (a) Comparison of the true and smoothed column densities for molecular clouds in the inner Milky Way; (b) Comparison of the true and distorted column densities for molecular clouds located in the upper-left region of the black solid line in panel (a); (c) Same as Fig.\ref{para}(g), but with black dashed and solid lines distinguishing molecular clouds affected by beam effects from those that are not; (d) Comparison of the velocity dispersion between molecular clouds in the inner Milky Way and those in nearby galaxies that are unaffected by beam effects; (e)
Comparison of cloud radii between molecular clouds in the inner Milky Way and those in nearby galaxies; (f) Comparison of cloud radii between molecular clouds in the inner Milky Way and those in nearby galaxies that are unaffected by beam effects.}
\label{origin}
\end{figure*}

\begin{figure}
\centering
\includegraphics[width=0.45\textwidth]{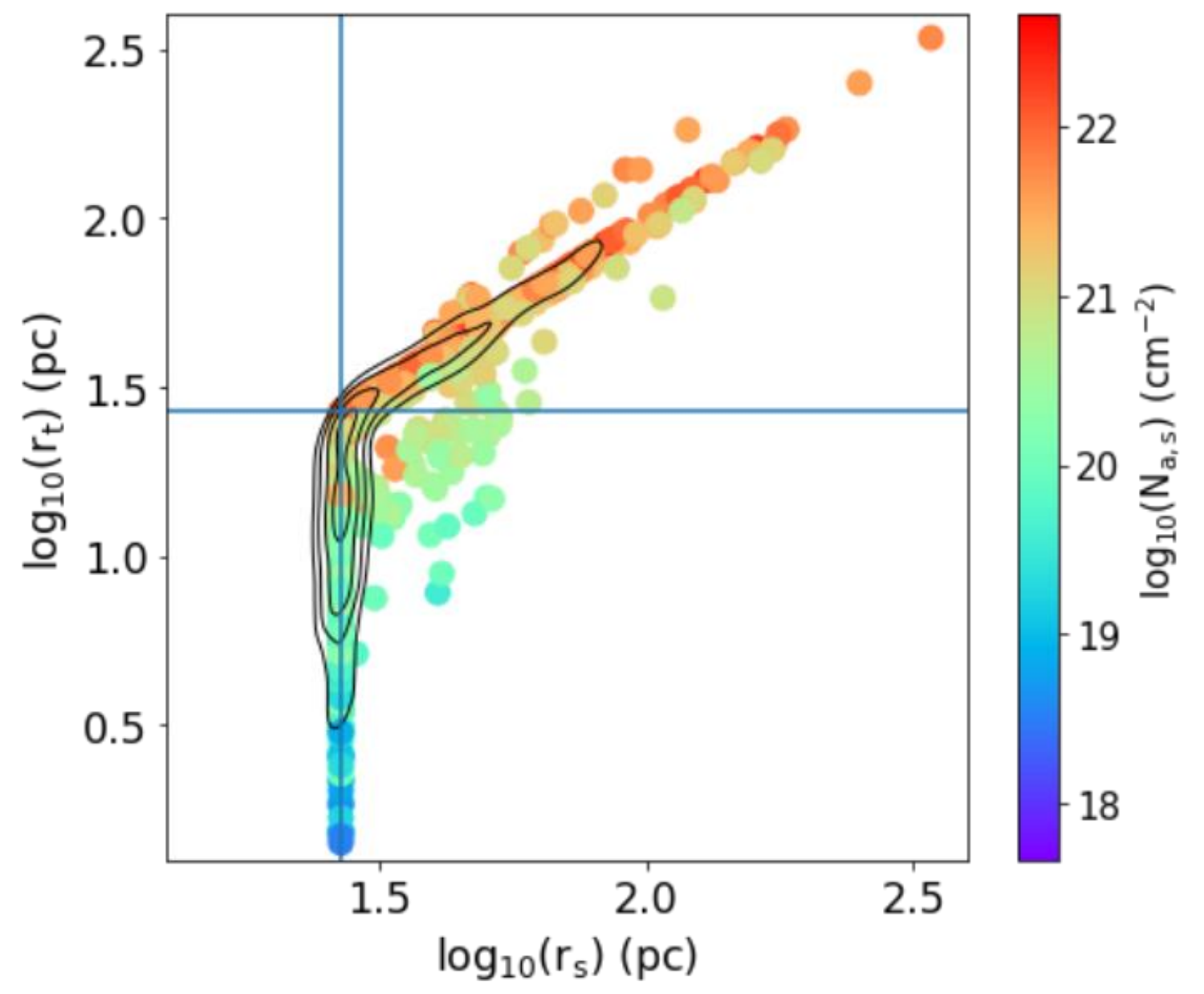}
\caption{Comparison of the real cloud radius ($r_{\rm t}$) and the smoothed cloud radius ($r_{\rm s}$), colored by $N_{\rm a,s}$. $N_{\rm a,s}$ is the smoothed average column density calculated based on $r_{\rm s}$. The vertical and horizontal lines mark the position at 53.5 pc. The black contours represent density levels of [0.3, 0.5, 0.7, 0.9] in the kernel density estimation (KDE) plot of the data points.}
\label{rn}
\end{figure}

The comparability of molecular clouds in the inner Milky Way and nearby galaxies, as presented above, is based on smoothing the Milky Way molecular clouds to an artificial scale. 
This highlights the need for caution when interpreting extragalactic observations, as the physical parameters of molecular clouds with sizes (diameters) comparable to or smaller than the beam size may be derived from artificial radius estimates. Consequently, these parameters may not accurately represent the true physical conditions of the clouds.
Below, we discuss the impact of resolution on the results by comparing the original high-resolution molecular clouds in the Milky Way with their smoothed counterparts. 

Smoothing does not affect mass and SFR estimates, as the total flux remains unchanged, but it does influence radius estimates, which in turn affect the estimation of density and other physical quantities. The estimate of a molecular cloud’s true radius ($r_{\rm t}$)  also depends on its filling factor ($f$).
Structures with a low filling factor ($f$ < 0.5) are usually more spatially concentrated in the image. This level of concentration was not important when the image was smoothed, as the structures were blurred. However, it becomes important now that we are estimating the true radius of molecular clouds at high resolution. The filling factor alone does not uniquely reflect the degree of concentration in the pixel distribution, as shown in Fig.\ref{true}. 
Therefore, we introduce an additional image parameter: the ratio of pixels in the largest connected region to the total number of non-zero pixels ($c$). A higher $c$ value indicates a more concentrated pixel distribution. The radius of the molecular cloud is ultimately determined according to the following rules:
(1) when the filling factor is high ($f \geq$ 2/3, Fig.\ref{true}(a)), $r_{\rm t} \sim r_{\rm max} \sim r_{\rm min}$; (2) for intermediate filling factors (1/3 < $f$ < 2/3, Fig.\ref{true}(b)), $r_{\rm t} \sim$  ($r_{\rm max}$+$r_{\rm min}$)/2; (3) For low filling factor ($f \leq$ 1/3), if $c$ > 0.5 (Fig.\ref{true}(c)), $r_{\rm t} \sim r_{\rm min}$. Note that 22 clouds ($\sim$2\% of the total) with $c$ < 0.5 are excluded because their radii are difficult to estimate (Fig.\ref{true}(d)). 


For the same cloud sample shown in Fig.\ref{para}, we now recalculate their physical parameters using $r_{\rm t}$ instead of $r_{\rm s}$.
In Fig.\ref{origin}(a), the column density derived using $r_{\rm t}$ by equation.\ref{N-a} is comparable to the original column density derived from equation.\ref{N-o}.
The strong correlation between column density and mass shown in Fig.\ref{para} is merely an artifact of image smoothing (the beam effect). As shown in Fig.\ref{origin}(a), there is no strong dependence between column density and mass. 

The beam effect can artificially enlarge the apparent size of molecular clouds, leading to inaccurate estimates of their radii. As a result, the derived densities of these clouds are also misleading.
In Fig.\ref{origin}(a), the straight green boundary line means the clouds on this line have a constant radius, i.e. $r_{\rm s}$ = 26.75 pc.
The densities of all molecular clouds located in the upper-left region of this line are distorted due to the beam effect. Fig.\ref{origin}(b) shows the true and distorted column densities. Apart from molecular clouds on the boundary line,
molecular clouds with column densities below the dashed black line ($N_{\rm 0} \sim 10^{20.9}$ cm$^{-2}$) are also significantly affected by the beam effect. This fact is more clearly shown in Fig.\ref{rn}.
For molecular clouds in three nearby galaxies, all of them are located in the lower right of the black solid line, but a subset lies below the black dashed line. These clouds correspond to those that significantly deviate from the mass-radius relation in Fig.\ref{origin}(e). 
As shown in Fig.\ref{origin}(f), after removing the molecular clouds affected by the beam effect, their mass-radius relation becomes comparable to that of molecular clouds in the inner Milky Way.
Here, we only consider the clouds of nearby galaxies located between the dashed and solid black lines in Fig.\ref{origin}(c).   
After this filter, in Fig.\ref{origin}(d), the velocity dispersions of molecular clouds in the inner Milky Way and nearby galaxies remain comparable.
However, a small fraction of molecular clouds in NGC 5236 exhibit higher velocity dispersions, as also shown in Fig.\ref{para}. This may reflect the influence of the galactic environment. As illustrated in Fig.\ref{sample}, NGC 5236 features a bar structure and strong spiral arms.

\subsection{Molecular clouds and their internal clumps}

\begin{figure*}
\centering
\includegraphics[width=1\textwidth]{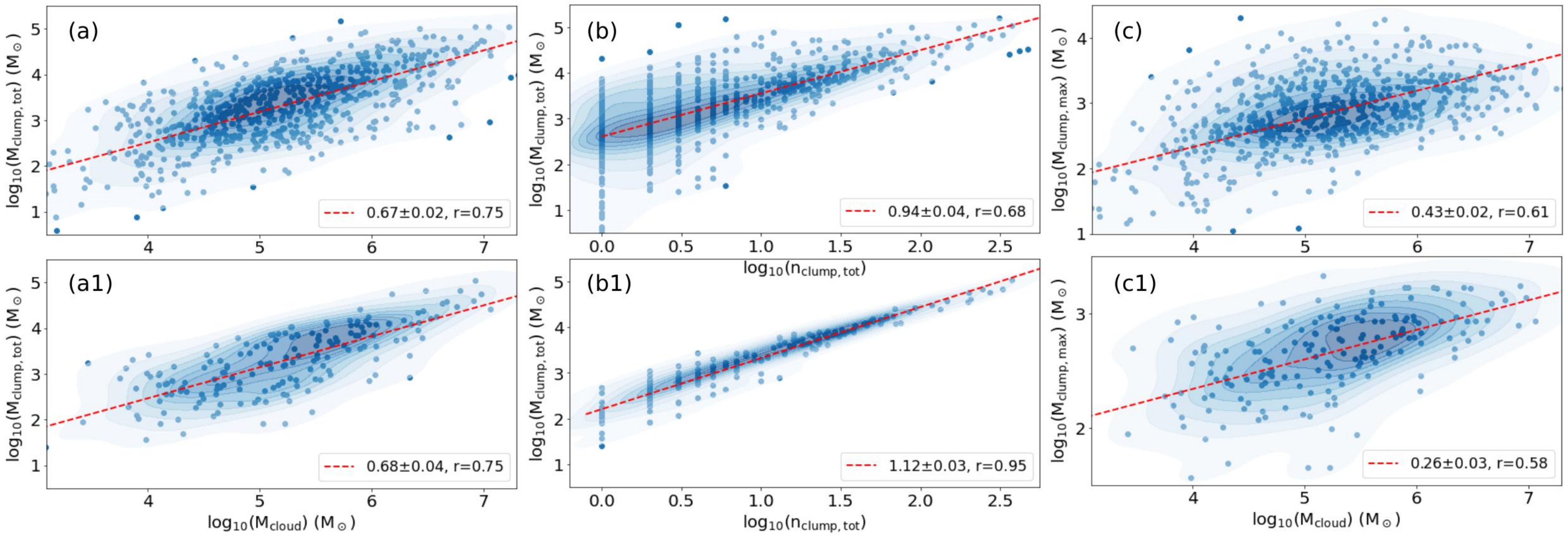}
\caption{Correlations between molecular clouds and their internal clumps under the upper limit case. (a) Correlation between the molecular cloud mass and the total clump mass within it;
(b) Correlation between the total clump mass and the number of clumps within the cloud;
(c) Correlation between the molecular cloud mass and the mass of its most massive internal clump.
The second row is the same as the first, but the molecular cloud distances are limited to within 2–4 kpc.}
\label{mtot}
\end{figure*}

\begin{figure}
\centering
\includegraphics[width=0.45\textwidth]{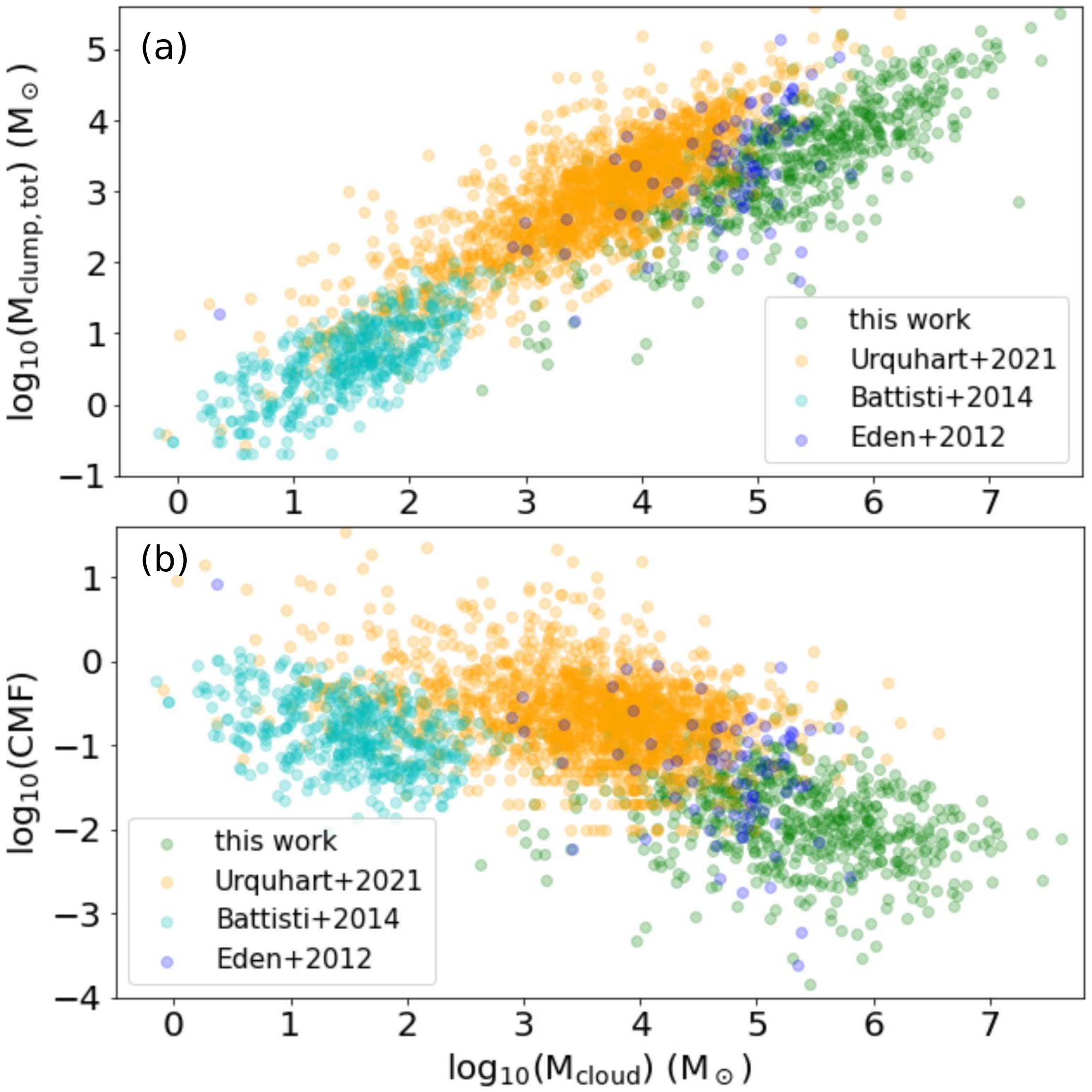}
\caption{Correlations between the total clump mass within a molecular cloud, the clump mass fraction and the cloud mass.}
\label{CMF}
\end{figure}

Star formation occurs primarily within clumps, the dense substructures of molecular clouds. In this work, the median values of the actual surface density and radius of the molecular clouds are 113 M$_\odot$/pc$^{2}$ and 25 pc, respectively, consistent with the statistical results of the molecular cloud catalog of \citet{Miville2017-834}. For the ATLASGAL clumps in the catalog of \citet{Urquhart2022-510}, these two values are 906 M$_\odot$/pc$^{2}$ and 0.32 pc, respectively.
In \citet{Urquhart2022-510}, the ATLASGAL clumps were grouped into complexes via friends-of-friends clustering, based on their spatial and velocity coherence. For all clump complexes, the median distance uncertainty is $\approx$0.5 kpc. Using the cloud distance determined in Sec.~\ref{inner} as a reference, clumps within a deviation of $r_{\rm t}$ (lower limit) or 0.5 kpc (upper limit) from it are considered to be physically associated with the molecular cloud. 
For the upper limit case, in Fig.\ref{mtot}(a), 
there is a clear positive correlation between the mass of a molecular cloud ($M_{\mathrm{cloud}}$) and the total mass of clumps ($M_{\mathrm{clump,tot}}$) within it. Although the number of resolvable clumps is affected by the distance of the cloud, Fig.\ref{mtot}(b) shows that the total clump mass and the total number of clumps within the cloud ($n_{\mathrm{clump,tot}}$) are also positively correlated. If we restrict the cloud sample to those within 2–4 kpc to avoid possible distance bias, as done in \citet{Urquhart2022-510}, a much clearer positive correlation can be observed in Fig.\ref{mtot}(b1).
For the lower limit case, these correlations persist, as shown in Fig.\ref{mtot1}.
We also attempted another method to match clumps with molecular clouds. Specifically, we extracted the averaged $^{13}$CO (1-0) spectrum based on each clump’s central coordinates and radius. If the velocity corresponding to the peak of the spectral line falls within the velocity range of a molecular cloud, the clump is considered to be associated with that cloud. We then assign the median distance of all associated clumps as the distance of the molecular cloud. Subsequently, all physical parameters of the clumps are corrected using this median distance. Comparing Fig.\ref{mtot2} and Fig.\ref{mtot}, this method strengthens the correlation between clumps and molecular clouds. The strongest correlations between $M_{\mathrm{cloud}}$, $M_{\mathrm{clump,tot}}$ and SFR for molecular clouds in the inner Milky Way derived in this work are
\begin{equation}
\mathrm{log}_{\rm 10} (M_{\mathrm{clump,tot}}) = 0.75 \times \mathrm{log}_{\rm 10} (M_{\mathrm{cloud}}) - 0.56 ,
\label{Mcto}
\end{equation}
\begin{equation}
\mathrm{log}_{\rm 10} (\mathrm{SFR}) = 0.98 \times \mathrm{log}_{\rm 10} (M_{\mathrm{cloud}}) - 9.12 .
\label{sfe-cloud}
\end{equation}

Differences in the star-forming ability of molecular clouds are reflected in the variation of their internal clump content. The correlation between a molecular cloud's physical properties and its star formation activity is also manifested through the presence and characteristics of its internal clumps.
Fig.\ref{mtot}(a) shows a significant correlation between the mass of molecular clouds and the total mass of their internal clumps. Similarly, Fig.\ref{mtot}(c) presents a clear correlation between the mass of a molecular cloud and that of its most massive internal clump ($M_{\mathrm{clump,max}}$). In addition, Fig.\ref{para} and Fig.\ref{origin} reveal strong correlations between cloud mass and other physical parameters, such as density, velocity dispersion, and virial parameter. These results suggest that the physical conditions of a molecular cloud may constrain the properties of its internal clumps, thereby influencing the cloud’s star-forming capability. 
The specific physical mechanisms and how they give rise to the correlations shown in Fig.\ref{mtot} require further investigation.

\citet{Zetterlund2019-881} matched Hi-GAL dust-identified 3674 clumps to 473 clouds traced by $^{12}$CO (3-2) emission, they also found that more massive clouds produce more clumps and more massive clumps. Moreover, they derived a mean clump mass fraction (CMF) of $\sim$0.2. The CMF is defined as the ratio of the total mass of clumps contained within a cloud to the total mass of the cloud (i.e. $M_{\mathrm{clump,tot}}$/$M_{\mathrm{cloud}}$), which is also called the dense gas fraction (DGF)
\citep{Urquhart2021-500,Eden2021-500}. 
\citet{Eden2012-422} and \citet{Battisti2014-780} matched BGPS clumps to molecular clouds traced by $^{13}$CO (1–0) emission and found that the CMF ranges from approximately 8\% to 13\%. Similarly, \citet{Urquhart2021-500} matched ATLASGAL clumps to molecular clouds traced by $^{13}$CO (2–1) emission and reported a CMF range of about 5\%–16\%. However, for our sample, the median CMF is only $\sim$1.4\%.
The catalogs of \citet{Eden2012-422,Battisti2014-780,Urquhart2021-500} are available online. Fig.\ref{CMF}(b) shows that the CMF is dependent on the molecular cloud mass. Firstly, different emission lines yield different estimates of molecular cloud masses, and the method used to identify molecular clouds can also affect the mass estimation. The molecular cloud masses identified in this work are significantly larger. To be precise, what is identified in this work are molecular cloud complexes. Previous studies, of course, were able to resolve the substructures within these complexes, which are then referred to as molecular clouds.
Moreover, \citet{Urquhart2021-500} used the ATLASGAL clump catalog of \citet{Urquhart2018-473}, which gives the full mass of the clump. However, the clump mass in the catalog of \citet{Urquhart2022-510} used in this work is determined by integrating the flux density at 870 $\mu$m within the Full Width at Half Maximum (FWHM) contour, i.e., above 50\% of the peak of the ATLASGAL dust continuum emission. The systematically lower clump masses also result in a smaller CMF. Overall, to compare CMFs from various works, the masses of molecular clouds and clumps must be measured consistently.

\subsection{Molecular clouds in the inner Milky Way and NGC 628}

\begin{figure*}
\centering
\includegraphics[width=0.75\textwidth]{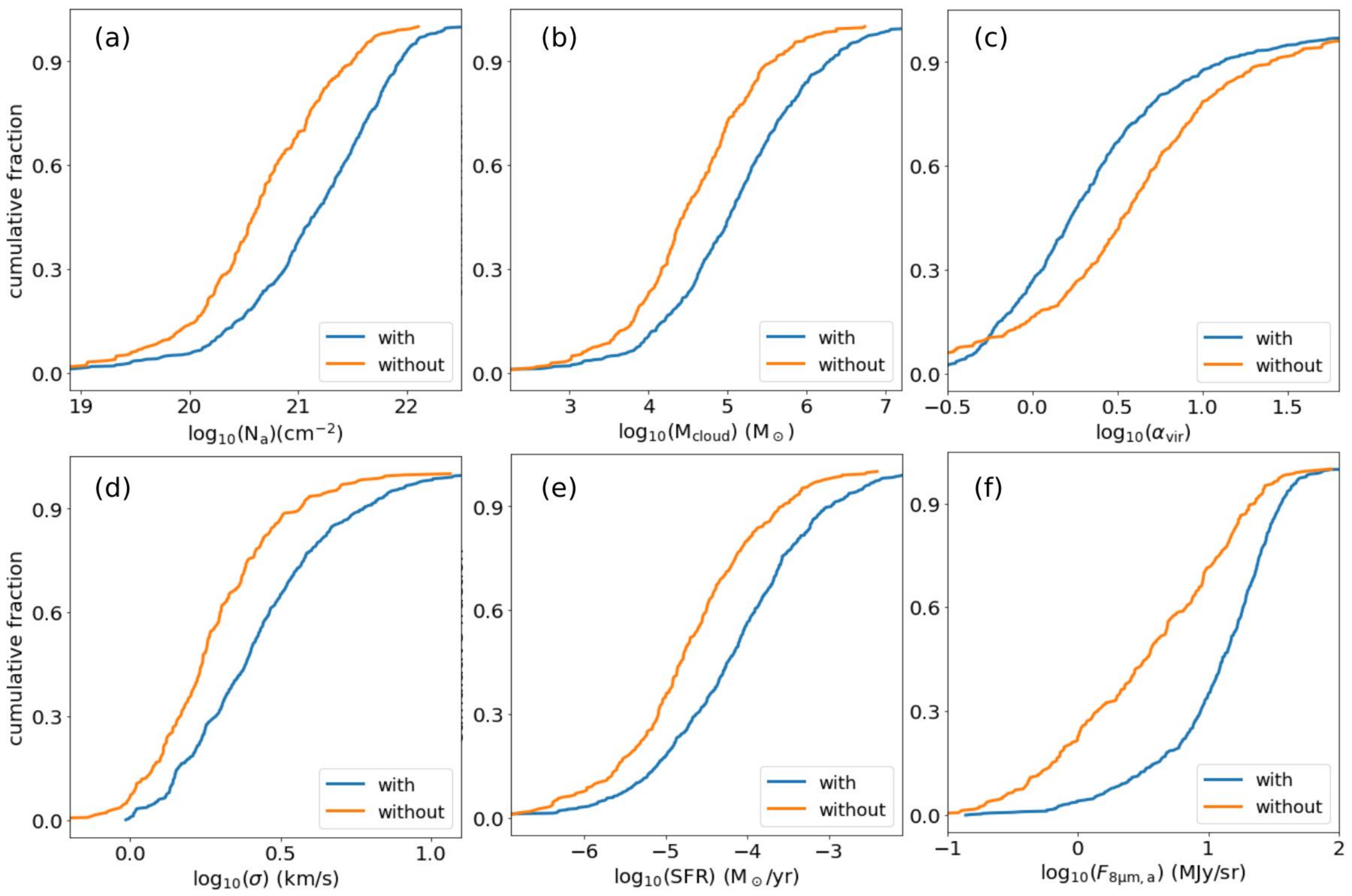}
\caption{Two groups of molecular clouds in the inner Milky Way, classified based on whether they are associated with clumps ('with' and 'without'). From panels (a) to (f), the physical quantities shown are the average column density, cloud mass, virial parameter, velocity dispersion, star formation rate, and average 8 $\mu$m surface brightness, respectively.}
\label{assoc}
\end{figure*}

In \citet{Zhou2024-534}, we conducted a detailed study of the molecular clouds in NGC 628, where the clouds were classified into three distinct categories (i.e. leaf-HFs-A, leaf-HFs-B and leaf-HFs-C), each exhibiting significantly different physical properties. leaf-HFs-A present the best hub-filament morphology, which also have the highest density contrast, the largest mass and the lowest virial ratio. The numbers of the associated 21 $\mu$m and H$_{\alpha}$ structures and the peak intensities of 7.7 $\mu$m, 21 $\mu$m and H$_{\alpha}$ emissions decrease from leaf-HFs-A to leaf-HFs-C. These evidence indicate that leaf-HFs-A have more active star formation than leaf-HFs-C. Since clumps are the real star-forming sites within molecular clouds, leaf-HFs-A, which have greater mass, contain more clumps and therefore exhibit stronger star formation activity.

As described in Sec.\ref{inner}, a considerable fraction of molecular clouds do not have associated clumps. Thus, we can roughly divide molecular clouds in the inner Milky Way into two groups based on whether they are associated with clumps ("with" and "without"). The group associated with clumps is clearly more active in star formation. In Fig.\ref{assoc}, molecular clouds in this group exhibit greater mass, higher density, larger velocity dispersion, and smaller virial parameters. These differences are also clearly reflected in the comparison between leaf-HFs-A and leaf-HFs-C molecular clouds in NGC 628.
As presented in \citet{Mazumdar2021-650,Zhou2024-682-128}, the 8 $\mu$m emission can be a good indicator of feedback strength. We calculated the average 8 $\mu$m surface brightness ($F_{\rm 8 \mu m,a}$) over each cloud to measure
the strength of feedback in each cloud. 
In Fig.\ref{assoc}(f), molecular clouds associated with clumps exhibit stronger 8 $\mu$m emission, which suggests more intense stellar feedback. Hence, their elevated velocity dispersion may be a consequence of star formation feedback.


\section{Conclusion}

We used the CO (2–1) and CO (1–0) data cubes to identify molecular clouds and investigate their gas kinematics and dynamics in three nearby galaxies and the inner Milky Way, respectively. We compare the physical properties of molecular clouds across these environments and find that, when observed at consistent resolution, their properties are broadly comparable. The star formation rates (SFRs) of molecular clouds in both the inner Milky Way and nearby galaxies also follow similar trends.

However, this comparability relies on smoothing Milky Way molecular clouds to an artificial scale. The beam effect can artificially enlarge the apparent size of molecular clouds, leading to inaccurate estimates of their radii, and consequently, misleading values for derived quantities such as density and virial parameter. By comparing the original high-resolution Milky Way clouds with their smoothed counterparts, we established criteria to exclude clouds significantly affected by the beam effect in the extragalactic sample.

After applying this filter, we re-examined the physical properties of molecular clouds in the inner Milky Way and the three nearby galaxies, and found that the properties remain broadly consistent. Nonetheless, a small fraction of molecular clouds in NGC 5236 exhibit higher velocity dispersions, which may reflect the influence of the galactic environment.

In the inner Milky Way, molecular clouds can be classified into two groups: those associated with clumps and those without. Clump-associated clouds tend to be more massive, denser, have higher velocity dispersions, and lower virial parameters. These clouds also show stronger 8 $\mu$m emission, suggesting more intense stellar feedback, which may contribute to their elevated velocity dispersions.

There is a strong positive correlation between a cloud’s mass and the total mass of clumps it contains. Likewise, the number of clumps scales with both the cloud’s mass and total clump mass. Additionally, a clear relationship exists between a cloud’s mass and that of its most massive clump. These findings indicate that a cloud’s physical conditions influence the properties of its internal clumps, thereby shaping its overall star-forming potential.

\section*{Acknowledgements}
Thanks to the referee for the detailed comments that helped clarify this work.
It is a pleasure to thank the PHANGS team, the data cubes and other data products shared by the team make this work can be carried out easily. 
ALMA is a partnership of ESO (representing its member states), NSF (USA) and NINS (Japan), together with NRC (Canada), NSTC and ASIAA (Taiwan), and KASI (Republic of Korea), in cooperation with the Republic of Chile. The Joint ALMA Observatory is operated by ESO, AUI/NRAO and NAOJ. 
This work is based on observations made with the NASA/ESA/CSA James Webb Space Telescope. 
This publication makes use of data from FUGIN, FOREST Unbiased Galactic plane Imaging survey with the Nobeyama 45-m telescope, a legacy project in the Nobeyama 45-m radio telescope.

\section{Data availability}
For nearby galaxies, all the data used in this work are available on the PHANGS team website
\footnote{\url{https://sites.google.com/view/phangs/home}}. For the inner Milky Way, the datasets are available on the team websites of the ThrUMMS \footnote{\url{https://gemelli.spacescience.org/~pbarnes/research/thrumms/rbank/}} and the FUGIN \footnote{\url{https://nro-fugin.github.io/release/}} surveys. 
The molecular cloud catalog produced in this work is available at the CDS.

\bibliography{ref}
\bibliographystyle{aasjournal}

\begin{appendix}

\section{Supplementary maps}

\begin{figure*}
\centering
\includegraphics[width=1\textwidth]{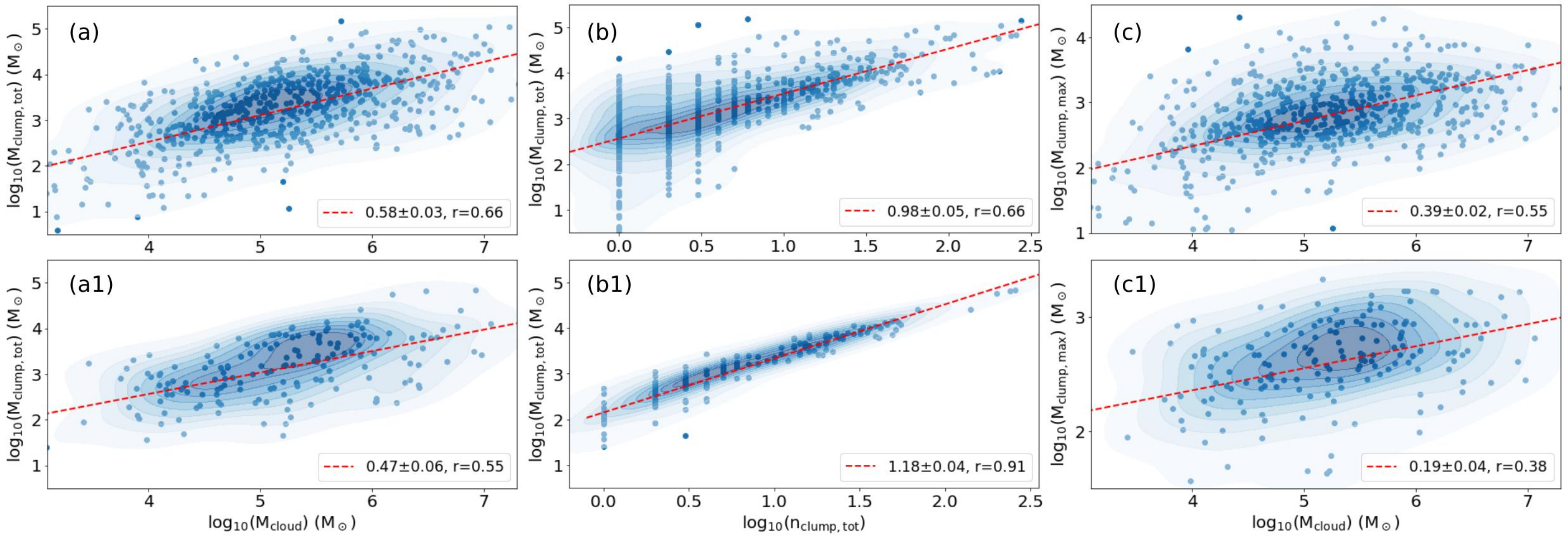}
\caption{Same as Fig.\ref{mtot}, but for the lower limit case.}
\label{mtot1}
\end{figure*}
\begin{figure*}
\centering
\includegraphics[width=1\textwidth]{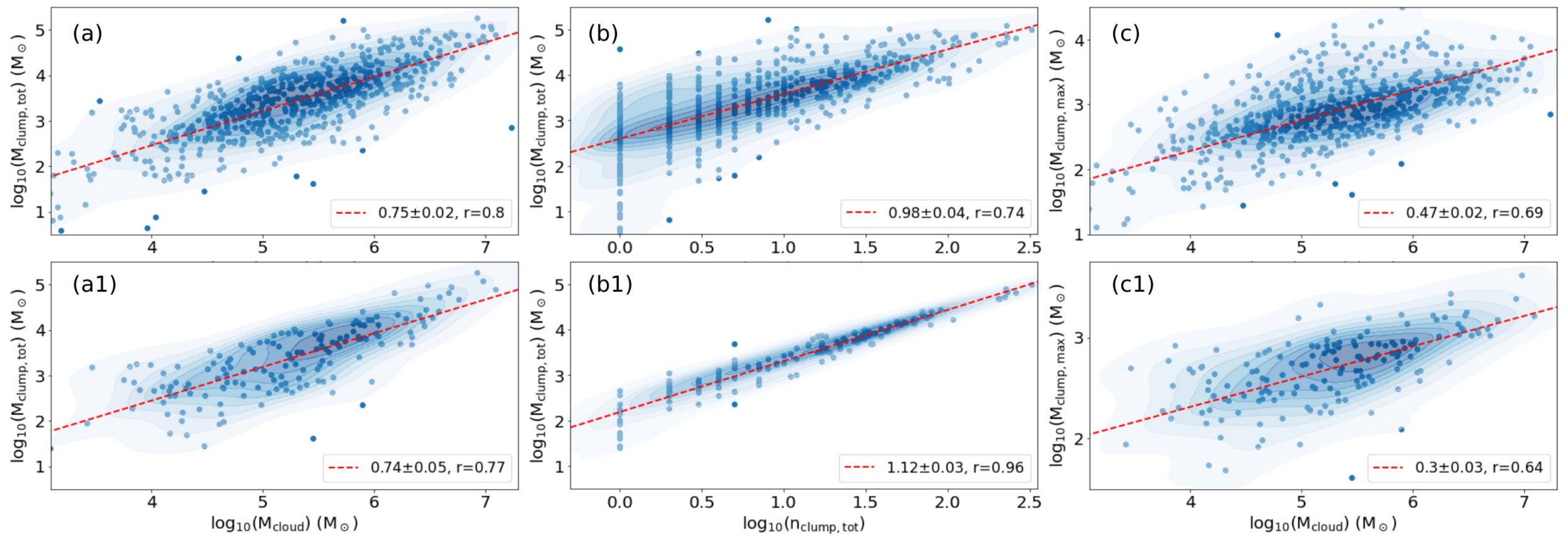}
\caption{Same as Fig.\ref{mtot}. The averaged $^{13}$CO (1–0) spectrum for each clump was obtained from its central coordinates and radius. When the velocity of the spectral peak coincided with the velocity range of a molecular cloud, the clump was identified as belonging to that cloud. The median distance of all matched clumps was taken as the molecular cloud’s distance, and the physical properties of the clumps were adjusted accordingly. This method is different from the one described in Sec.\ref{inner}.}
\label{mtot2}
\end{figure*}


\end{appendix}

\end{document}